\documentclass[usenatbib]{mnras}
\usepackage{newtxtext,newtxmath}
\usepackage[T1]{fontenc}
\usepackage{ae,aecompl}

\usepackage{graphicx}	
\usepackage{amsmath}	
\usepackage{amssymb}	
\usepackage{ulem}

\title{A system of three transiting super-Earths in a cool dwarf star}

\author[E.D. Alonso et al.]{E. D\'iez Alonso,$^{1}$
S. L. Su\'arez G\'omez,$^{1}$
J. I. Gonz\'alez Hern\'andez,$^{2,3}$
\newauthor A. Su\'arez Mascare\~no,$^{4}$
C. Gonz\'alez Guti\'errez$^{1},$
S. Velasco$^{2,3},$
B. Toledo--Padr\'on$^{2,3},$
\newauthor F. J. de Cos Juez$^{1}$\thanks{E-mail: fjcos@uniovi.es}, R. Rebolo$^{2,3,5}$
\\
$^{1}$Department of Exploitation and Exploration of Mines, University of Oviedo, Oviedo, Spain\\
$^{2}$Instituto de Astrof\'isica de Canarias, E--38205 La Laguna, Tenerife, Spain\\
$^{3}$Universidad de La Laguna, Dpto. Astrof\'isica, E--38206 La Laguna, Tenerife, Spain\\
$^{4}$Observatoire Astronomique de l\textquotesingle Universit\'e de G\`eneve, 1290 Versoix, Switzerland\\
$^{5}$Consejo Superior de Investigaciones Cient\'ificas, Spain\\
}

\date{Accepted XXX. Received YYY; in original form ZZZ}

\pubyear{2017}

\begin{document}
\label{firstpage}
\pagerange{\pageref{firstpage}--\pageref{lastpage}}
\maketitle

\begin{abstract}

We present the detection of three super-Earths transiting the cool star LP415-17, monitored by K2 mission in its $13^{th}$ campaign. High resolution spectra obtained with HARPS-N/TNG showed that the star is a mid-late K dwarf. Using spectral synthesis models we  infer its effective temperature, surface gravity and  metallicity and subsequently determined  from evolutionary models a stellar radius of 0.58 $R_{\odot}$. The planets have radii of 1.8, 2.6 and 1.9 $R_{\oplus}$ and orbital periods of 6.34, 13.85 and 40.72 days. High resolution images discard any significant contamination by an intervening star in the line of sight. The orbit of the furthest planet has radius of 0.18 AU, close to the inner edge of the habitable zone. The system is suitable to improve our understanding of formation and dynamical evolution of super-Earth systems in the rocky -- gaseous threshold, their atmospheres, internal structure, composition and interactions with host stars.

\end{abstract}




\section{INTRODUCTION}
Transiting planetary systems are of great value  for the characterization of exoplanetary atmospheres \citep{2000ApJ...529L..45C,2014Natur.505...69K}, the understanding of planetary formation and evolution \citep{1999Natur.402..269O} and for the study of the interactions with their host stars \citep{2017AJ....153..185C}. Cool dwarf stars are well suited to find Earth size and super-Earth size planets by the transit method, since transits produce  deeper dimmings than in solar-type stars. The amplitude of signals in transit transmission spectroscopy is also higher for stars with small radius, favoring these type of stars for future atmospheric studies of their planetary systems \citep{2016Natur.533..221G}.

The Kepler mission \citep{2010Sci...327..977B} has achieved a large number of detections of transiting planets \citep{2012ApJS..201...15H}. In its second mission \citep{2014PASP..126..398H}, the satellite performs observations of different ecliptic plane fields for periods of time spanning about 80 days. Multiple signals of exoplanet candidates are present in concluded campaigns \citep{2015ApJ...800...59V,2016ApJS..226....7C}. Campaign $13^{th}$ has focused in the Hyades and Taurus region, centered in $\alpha$=04:51:11, $\delta$=+20:47:11, between 2017 March $08^{th}$ and 2017 May $27^{th}$. In this campaign K2 observed 26.242 targets at standard long cadence mode and 118 targets at short cadence mode. 
LP415-17 ($\alpha$=04:21:52.487, $\delta$=+21:21:12.96) has been observed in low cadence mode.

In this letter, we present the detection of three super-Earths transiting the star LP415-17 (EPIC 210897587, 2MASS 04215245+2121131). In section 2 we describe the characterization of the star from spectra acquired with HARPS-N spectrograph and the analysis of the K2 photometric time series. We also discuss possible contaminating sources and the main parameters derived for the planets. In section 3 we present estimations of the masses, discuss the stability of the system and its suitability for future characterization by transmission spectroscopy. In section 4 we summarize the main conclusions derived from this work.

\section{METHODS}

\subsection{Stellar characterization}

We obtained three spectra, 1800 seconds of exposure time each, with HARPS-N \citep{2012SPIE.8446E..1VC} a fibrefed high resolution echelle spectrograph installed at the 3.6 m  Telescopio Nazionale Galileo in the Roque de los Muchachos Observatory (Spain) with  a resolving power of R = 115,000 over a spectral range from  380 to 690 nm.
We have averaged them to obtain a final spectrum smoothed with 10-pixel step to improve the visualization of spectral features.
We compared this spectrum with synthetic models \citep{2014A&A...568A...7A} generated with ASSET \citep{2008ApJ...680..764K}, adopting Kurucz atmospheric models \citep{2003IAUS..210P.A20C,2012AJ....144..120M}.
Synthetic spectra have been broadened with a macroturbulence profile of 1.64 km/s \citep{2005ApJ...622.1102F}, with a rotation profile taking $v_{rot}$ = 1.8 km/s \cite{2005oasp.book.....G} and with a gaussian profile for instrumental broadening at resolution of 115,000 (FWHM~$\sim 2.6$~km/s).
The effective temperature has been estimated using the infrared flux method \citep{2009A&A...497..497G}. We apply $T_{\rm IRFM}$ -- (color, [Fe/H]) calibrations for dwarf stars, correcting for extinction the stellar magnitudes according to $A_{X}=R_{X}\times E(B-V)$. $R_{X}$ was obtained from \citet{2004AJ....128.2144M}, and the reddening E(B-V) from the dust maps \citep{1998ApJ...500..525S}, but corrected using the equations in \cite{2000AJ....120.2065B} for the estimated distance to the star (82 pc) and galactic latitude. We obtained E(B-V)=0.087 and $T_{\rm IRFM}$= 4258 $\pm$ 150 K.

\begin{figure*}
	
	\includegraphics[width=0.8\textwidth] {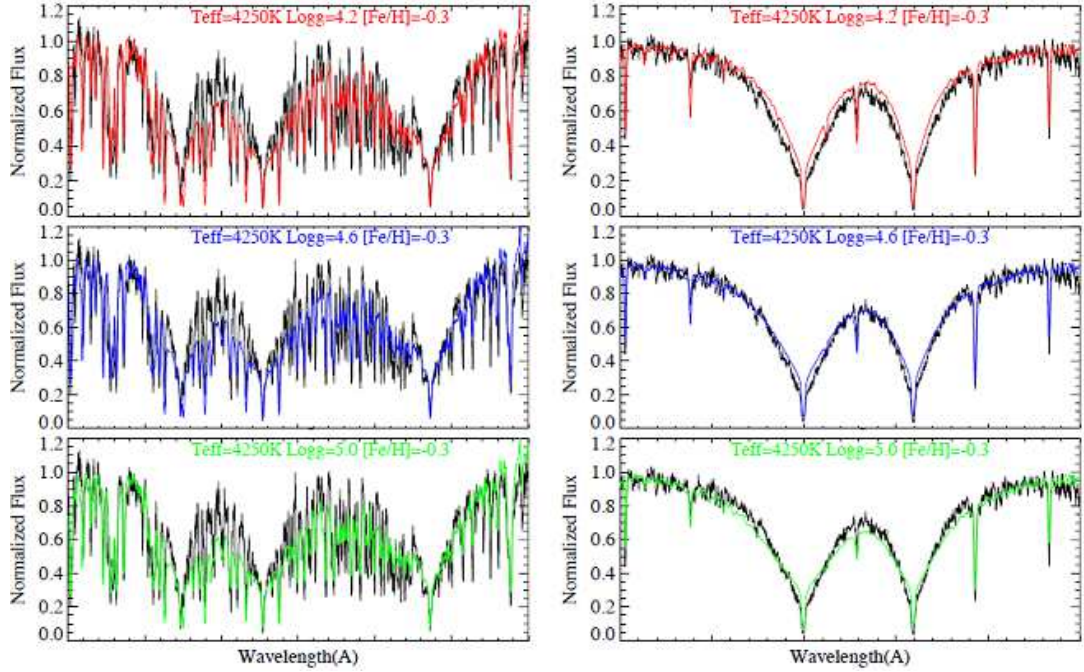}
    \caption{Synthetic spectral models centered in the magnesium triplet (left) and Na D doublet (right), calculated with $T_{\rm eff}$=4250 K, [Fe/H]=-0.3 and $\log g=3.4$ (top), $\log g=4.6$ (middle) and $\log g=4.0$ (bottom), compared with the observed spectrum. The best fit is obtained with $\log g=4.6$.}
    \label{fig:epic_spec}
\end{figure*}

\begin{figure}
	
	\includegraphics [width=0.5\textwidth]{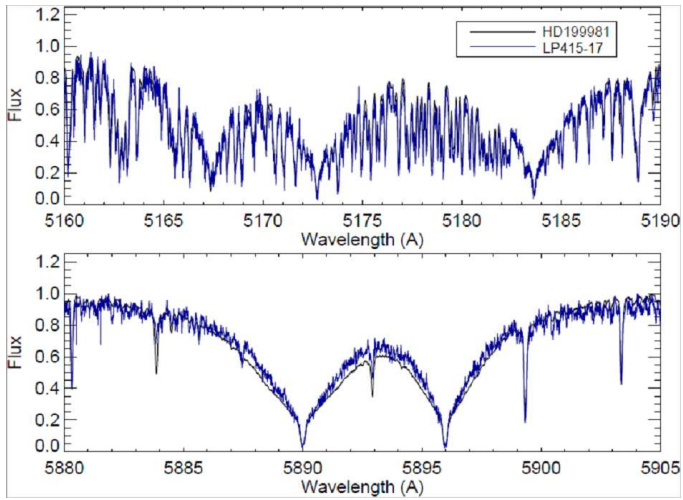}
    \caption{Spectrum of LP415-17 compared with  the observed spectrum of HD 199981.}
    \label{fig:HD199981}
\end{figure}

The metallicity and $\log g$ have been obtained via comparison of the observed spectrum with synthetic spectra, resulting [Fe/H] = $-0.3\pm0.2$ and $\log g=4.6\pm0.3$ (Fig.~\ref{fig:epic_spec}).
We compare the stellar parameters and metallicity with a grid of tabulated isochrones \citep{1994A&AS..106..275B}, and obtain $R_{*}$=$0.58^{+0.06}_{-0.03}$ $R_{\odot}$, $M_{*}$=$0.65^{+0.06}_{-0.03}$ $M_{\odot}$ and $M_{V}$=$7.95^{+0.34}_{-0.66}$ mag.
In Fig.~\ref{fig:HD199981} we compare the observed spectrum of LP415-17 with that of the well characterized star HD199981 \citep{2013AJ....146..134K}, whose parameters (K6V, $T_{\rm eff}=4263K$, $\log g=4.97$) are very close to our results for LP415-17.

Adopting $m_{V}$ = 12.54, we estimate a distance to LP415-17 of $D_{*}$=$82^{+29}_{-12}$ pc. We measured a radial velocity from the HARPS-N spectrum,  which combined with the proper motions reported in Table ~\ref{tab:stellar_parameters} results in velocity components U=-38.2 km/s, V=-69.4 km/s, W=31.8 km/s.  A comparison with the kinematic and metallicity properties of the Copenhaguen Survey of the Solar neighbourhood \citep{2004A&A...418..989N}, suggests that  LP 415-17  could be a member of the  Hercules stream, which according to \cite{2007ApJ...655L..89B} could be  a combination of thin and thick disk stars originating in interactions of the inner disk with  the bar of our Galaxy.

From the observed spectrum we determine a $R'_{HK}$  index of -4.85 $\pm$ 0.13 which according to \cite{2015MNRAS.452.2745S} indicates a likely rotation period of 34.8 $\pm$ 8.2 days. Table ~\ref{tab:stellar_parameters} summarizes stellar parameters for LP415-17.

\begin{table}
	  
	\centering
	\caption{Stellar parameters for LP415-17.}
	\label{tab:stellar_parameters}
    \resizebox{6 cm}{!} {
	\begin{tabular}{lcc}
		       
        Parameter&Value&Source\\
        \hline
        \hline
        V [mag]& $12.806 \pm 0.005$&(1)\\
        R [mag]& $12.286 \pm 0.006$&(1)\\
        I [mag]& $12.289 \pm 0.090$&(1)\\
        J [mag]& $10.274 \pm 0.021$&(2)\\
        H [mag]& $9.686 \pm 0.021$&(2)\\
        K [mag]& $9.496 \pm 0.014$&(2)\\
        $T_{\rm eff}$ [K]& 4258 $\pm$ 150&(3)\\
        $[Fe/H]$& $-0.3\pm0.2$&(3)\\
       	Radius [$R_{\odot}$] &$0.58^{+0.06}_{-0.03}$&(3)\\
		Mass [$M_{\odot}$]& $0.65^{+0.06}_{-0.03}$&(3)\\
        $\log g$ [cgs]& $4.6\pm0.3$&(3)\\
        $M_{V}$ [mag]& $7.95^{+0.34}_{-0.66}$&(3)\\
        $\log_{10} (R'_{HK})$& -4.85 $\pm$ 0.13&(3)\\
        $P_{rot}$ [d]&$34.8\pm8.2$&(3)\\
        Distance [pc] & $82^{+29}_{-12}$&(3)\\
        $V_{r}$ [km/s]&$19.1\pm 0.5$&(3)\\
        $\mu_{\alpha}$ [mas/y]&$201.9 \pm 6.9$&(1)\\
        $\mu_{\delta}$ [mas/y]&$-71.3 \pm 4.3$&(1)\\
        U, V, W  [km/s]&-38.2, -69.4, 31.8&(3)\\
       &&\\
        \multicolumn{3}{l}{(1) UCAC4 \citep{2013AJ....145...44Z}.}\\
         \multicolumn{3}{l}{(2) 2MASS \citep{2003tmc..book.....C}.}\\
          \multicolumn{3}{l}{(3) This work.}\\
            
	\end{tabular}
}
\end{table}

\subsection{K2 photometric data}

\begin{figure}
	\includegraphics [width=0.5\textwidth]{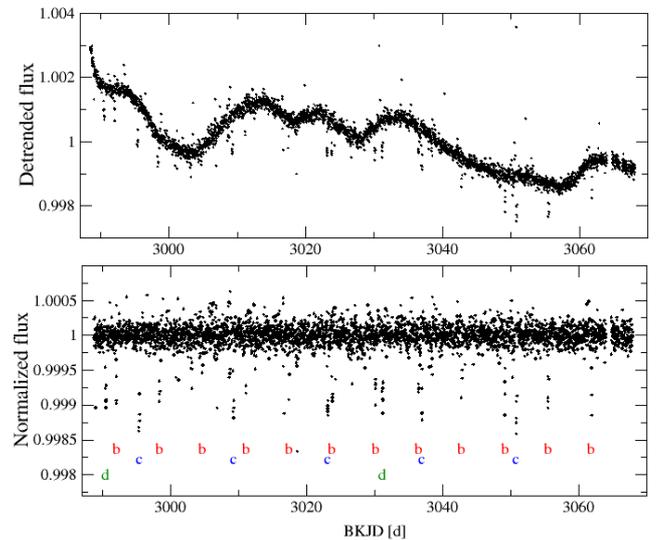}
    \caption{Top: K2 detrended light curve for LP415-17. Bottom: normalized light curve. Characters b, c and d indicate times of observed transits of planets b, c and d.}
    \label{fig:lightcurves}
\end{figure}

\begin{figure}
	\includegraphics [width=0.45\textwidth]{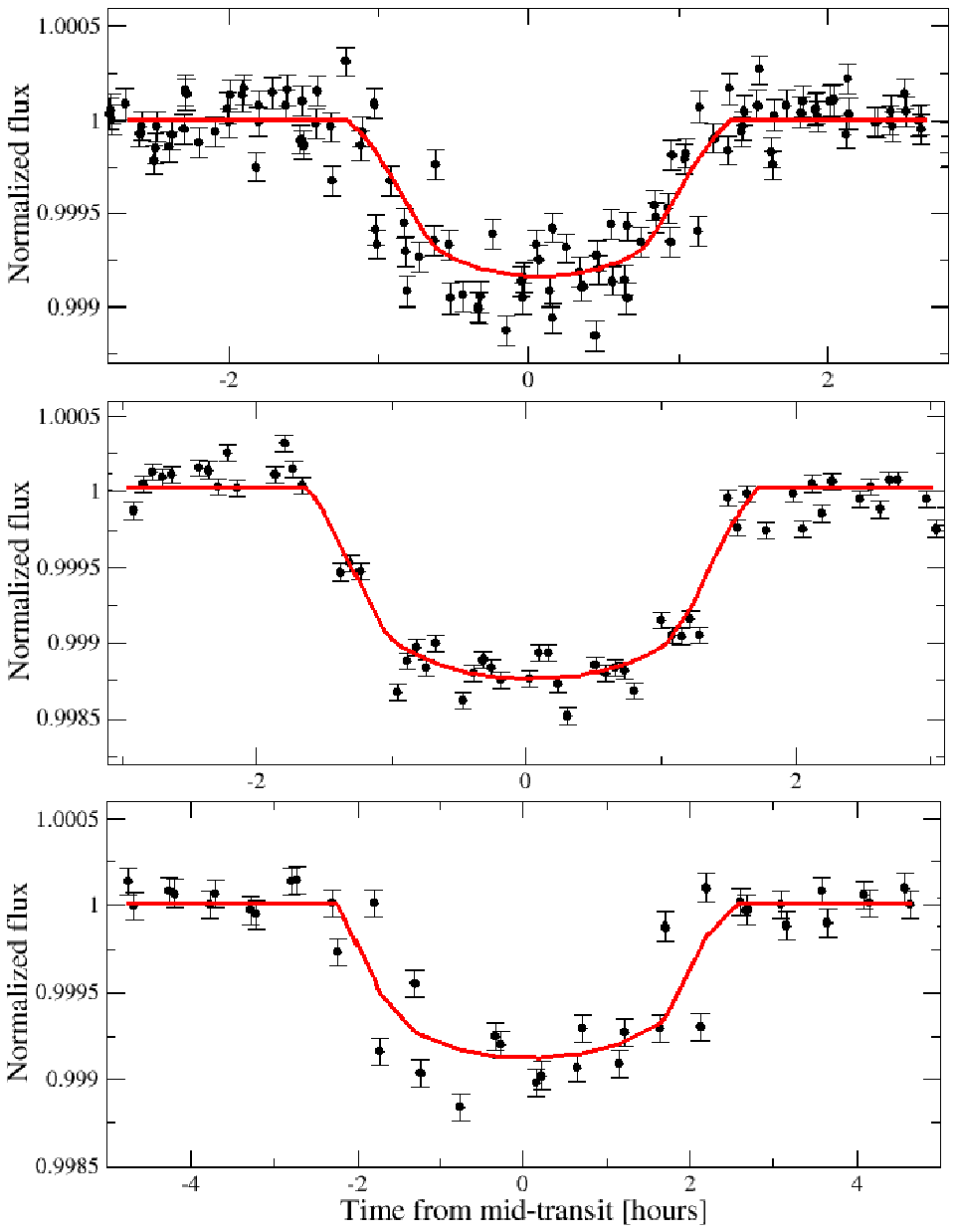}
    \caption{Phase-folded light curves corresponding to planets b (top), c (middle), and d (bottom). Solid curves represent best model fits obtained by MCMC.}
    \label{fig:transits}
\end{figure}

The light curve of LP415-17 exhibits clear signals of at least three transiting objects (Fig.~\ref{fig:lightcurves}). We analyzed the K2 corrected photometry from the star following the work of \cite{2014PASP..126..948V}, applying a spline fit to detrend stellar variability and search for periodic signals with a Box Least Squares (BLS) method \citep{2002A&A...391..369K} on flattened data. Once BLS finds a transit signal, it is fitted, removed and another search for transit signals is performed. Following this method, we find three transit signals of planet candidates with orbital periods 6.342$\pm$0.002 days (b), 13.850$\pm$0.006 days (c) and 40.718$\pm$0.005 days (d).

To estimate the main parameters for each planet we analyzed each phase-folded transit using MCMC, fitting models from \cite{2002ApJ...580L.171M} with the EXOFAST package \citep{2013PASP..125...83E}, resampling the light curve 10 times uniformly spaced over the 29.4 minutes for each data point and averaging \citep{2010MNRAS.408.1758K} (Fig.~\ref{fig:transits}). We set priors on host star ($T_{\rm eff}$=4258K, $\log g=4.6$, [Fe/H]=-0.3), orbital periods ($P_{orb}$= 6.34 (b), 13.85 (c), 40.72 (d)) days. Due to tidal circularization, e=0 for planet b. We also assume e=0 for c and d, for beeing transiting planets in multi planetary systems \citep{2015ApJ...808..126V}. The parameters obtained for planets b, c and d are summarized in Table ~\ref{tab:planetparams}.

Planet d only shows two transits in the K2 $13^{th}$ campaign observation window. Separate analysis of these two transits reveals coincident transit parameters, supporting the idea that this signal is created by the same object. The parameters for each observed transit of planet d are summarized in Table ~\ref{tab:planetparams}.

The planets have estimated radii $1.8^{+0.2}_{-0.1}$ $R_{\oplus}$ (b), $2.6^{+0.7}_{-0.2}$ $R_{\oplus}$ (c) and $1.9^{+0.7}_{-0.2}$ $R_{\oplus}$ (d), orbital periods 
6.342$\pm$0.002 days (b), 13.850$\pm$0.006 days (c) and 40.718$\pm$0.005 days (d), and semimajor axis $0.0562^{+0.0013}_{-0.0014}$ AU (b), $0.0946^{+0.0031}_{-0.0030}$ AU (c) and $0.1937^{+0.0064}_{-0.0059}$ AU (d).

\subsection{False positives analysis}

\begin{figure}
	
	\includegraphics [width=0.4\textwidth]{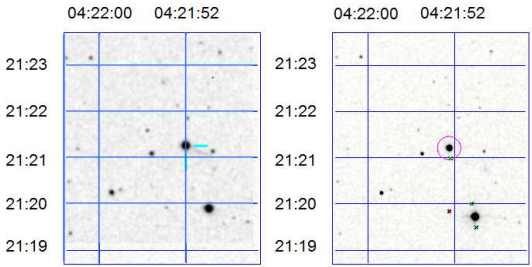}
    \caption{POSS-I image (1953, left) and 2MASS (1998, right). No background source is present at star's position in K2 $13^{th}$ campaign.}
    \label{fig:POSS-I_2MASS}
\end{figure}

To exclude false positives from possible companions, we analyzed speckle images of the star at 562 and 832 nm, obtained with NESSI at the 3.5 meters WIYN telescope (Kitt peak, Arizona), available at ExoFOP--K2 \footnote{https://exofop.ipac.caltech.edu/k2/} \footnote{Images where taken by \cite{2018arXiv180106957H}. We noticed their work in this system during the revision process.}. Images exclude companions at 0.2 arcseconds with $\delta$mag < 3.5 and at 1 arcsecond with $\delta$mag < 6. We searched for possible contaminating background sources in images from POSS-I \citep{1963bad..book..481M} (year 1953) and 2MASS \citep{2003tmc..book.....C} (year 1998). LP415-17 exhibits high proper motion $\mu_{\alpha}$=201.9 mas/year  $\mu_{\delta}$=-71.3 mas/year so we can inspect for background sources at LP415-17's position during the K2 $13^{th}$ campaign. No background object is found at the current star position (Fig.~\ref{fig:POSS-I_2MASS}).

Speckle images from WIYN, inspection of POSS-I and 2MASS images, statistical analysis performed with vespa package \citep{2012ApJ...761....6M,2015ascl.soft03011M} (obtaining FPP < $10^{-5}$ for all the planets) and taking into account FPP overestimation for multi planet systems \citep{2011ApJS..197....8L,2016ApJ...827...78S}, make us to reject possible contaminating sources, concluding that the transit signals in LP 415-17 are of planetary origin.

\section{Discussion}

Following the mass -- radius relation \citep{2014ApJ...783L...6W}
\[\frac{M_{p}}{M_{\oplus}} = 2.69\cdot \Big(\frac{R_{p}}{R_{\oplus}}\Big)^{0.93}\]

for planets satisfying 1.5 $\leq$ $R_{p}$/$R_{\oplus}$ $\leq$ 4, we obtain $M_{b}$=4.7 $M_{\oplus}$, $M_{c}$=6.5 $M_{\oplus}$, $M_{d}$=4.9 $M_{\oplus}$ for planets b, c, and d, respectively. Assuming $M_{p} \ll M_{*}$, circular orbits and $\sin i\sim$1, we compute induced amplitudes in stellar velocity variations of 2.2 m/s for planet b, 2.3 m/s for planet c and 1.2 m/s for planet d.

Given the visual magnitude of the star (V=12.8)  the required  radial velocity monitoring  at 1 m/s precision is hard to obtain for HARPS-like spectrographs in medium size telescopes, however is well suited  for ESPRESSO \citep{2014AN....335....8P,2017arXiv171105250G} at the VLT. The moderated chromospheric activity of the star is likely to induce RV signals of order less than 3 m/s \citep{2017MNRAS.468.4772S} which should  not prevent the detection of the dynamical signals induced by the planets and the determination of their masses and densities. We estimate the incident flux for planet d as $F_{p}$=2.63 $F_{\oplus}$. Habitable zone estimations \citep{2016ApJ...830....1K} place the inner edge of the habitable zone at 1.5 $F_{\oplus}$, so planets b, c and d are closer to the star than the inner part of the habitable zone.

\begin{table*}
	\centering
   
	\caption{Parameters for planets b, c and d.}
	\label{tab:planetparams}
	
    \resizebox{17cm}{4.8cm} {
    \begin{tabular}{lccccc} 
    	
        Planet Parameters&b&c&d&\\
        \hline
        \hline
        Orbital period (P) [d]& 6.342$\pm$0.002&13.850$\pm$0.006&40.718$\pm$0.005&&\\
         \hline
        Semi-major axis (a) [AU]& $0.0562^{+0.0013}_{-0.0014}$&$0.0946^{+0.0031}_{-0.0030}$&$0.1937^{+0.0064}_{-0.0059}$&&\\
         \hline
        
Radius ($R_{p}$) [$R_{\oplus}$] & $1.8^{+0.2}_{-0.1}$&$2.6^{+0.7}_{-0.2}$&$1.9^{+0.7}_{-0.2}$&&\\
		 \hline
        Mass ($M_{p}$) [$M_{\oplus}$] (1)  & $4.7^{+0.5}_{-0.3}$&$6.5^{+1.5}_{-0.5}$&$4.9^{+1.7}_{-0.6}$&&\\
        \hline
        
        Equilibrium Temperature ($T_{eq}$) [K] & $708^{+38}_{-31}$&$583^{+52}_{-35}$&$381^{+47}_{-25}$&&\\
        
        &&&&&\\
 
   		Transit Parameters& b&c&d& d (2990.500) (2)& d (3031.218) (2)\\
        \hline
        \hline
       
        Number of transits& 12&5&2&--&--\\
        \hline
        Epoch (BKJD) [days]& 2992.068$\pm$0.002&2995.426$\pm$0.001&2990.500$\pm$0.003&2990.500$\pm$0.003&3031.218$\pm$0.002\\
        \hline
 
        Radius of planet in stellar radii ($R_{p}$/$R_{*}$)&$0.0261^{+0.0009}_{-0.0008}$ &$0.0321^{+0.0016}_{-0.0010}$&$0.0273^{+0.0021}_{-0.0015}$&$0.0280^{+0.0011}_{-0.0009}$ &$0.0292^{+0.0027}_{-0.0018}$\\
        \hline

        Semi major axis in stellar radii         (a/$R_{*}$)&$18.5.0^{+1.1}_{-1.6}$ &$27.0^{+2.2}_{-4.2}$&$63.3^{+6.4}_{-13}$&$73.2^{+4.2}_{-6.7}$ &$75.0^{+13}_{-20}$ \\
        \hline
    
        Linear limb-darkening coeff ($u_{1}$)&$0.644^{+0.063}_{-0.080}$ &$0.614^{+0.065}_{-0.084}$&$0.597^{+0.085}_{-0.110}$&$0.551^{+0.083}_{-0.11}$ &$0.540^{+0.10}_{-0.12}$\\
        \hline
        
        Quadratic limb-darkening coeff ($u_{2}$)&$0.122^{+0.074}_{-0.061}$ &$0.148^{+0.071}_{-0.060}$&$0.142^{+0.091}_{- 0.072}$&$0.174^{+0.093}_{-0.077}$ &$0.183^{+0.11}_{-0.088}$\\
        \hline
               
        Inclination (i) [deg]&$88.3^{+1.2}_{-1.9}$ &$88.96^{+0.71}_{-0.88}$&$89.61^{+0.27}_{-0.48}$&$89.75^{+0.17}_{-0.20}$ &$89.62^{+0.26}_{-0.37}$\\
        \hline 
       
        Impact Parameter (b)&$0.34^{+0.24}_{-0.23}$ &$0.45^{+0.37}_{-0.31}$&$0.44^{+0.33}_{-0.30}$&$0.33^{+0.20}_{-0.21}$ &$0.50^{+0.22}_{-0.31}$\\     
        \hline
        
        Transit depth ($\delta$)&$0.00068^{+0.00005}_{-0.00004}$ &$0.00103^{+0.00010}_{-0.00006}$&$0.00074^{+0.00012}_{-0.00008}$&$0.00078^{+0.00006}_{-0.00005}$ &$0.00085^{+0.00017}_{-0.00010}$\\
        \hline
                       
        Total duration ($T_{14}$) [days]&$0.1004^{+0.0050}_{-0.0048}$&$0.1470^{+0.0130}_{-0.0260}$&$0.1860^{+0.0097}_{-0.0018}$&$0.1713^{+0.0056}_{-0.0051}$&$0.156^{+0.018}_{-0.011}$\\
        \hline     
        
       Ingress/egress duration ($\tau$) [days]&$0.0030^{+0.0009}_{-0.0003}$ &$0.0056^{+0.0039}_{-0.0008}$&$0.0060^{+0.0050 }_{-0.0010}$&$0.0052^{+0.0014}_{-0.0005}$ &$0.0058^{+0.0049}_{-0.0016}$\\

        &&\\
        \multicolumn{6}{l}{(1): The masses are estimated using mass-radius relation from Weiss \& Marcy (2014).}\\
        \multicolumn{6}{l}{(2): Derived parameters for individual transits of planet d.}\\
       
	\end{tabular}

}

\end{table*}

The amplitude of the signal in transit transmission spectroscopy can be estimated as $\frac{R_{p} \cdot {h_{\rm eff}}}{(R_{*})^{2}}$ \citep{2016Natur.533..221G} with $h_{\rm eff}$ the effective atmospheric height, wich is related to the atmospheric scale height H = K$\cdot$T/$\mu$$\cdot$g (K Boltzmann's constant, T atmospheric temperature, $\mu$ atmospheric mean molecular mass, g surface gravity). We adopt $h_{\rm eff}$ = 7$\cdot$H \citep{2010ApJ...716L..74M} for a transparent volatile dominated atmosphere ($\mu$ = 20) with 0.3 Bond albedo. With these assumptions we estimate the amplitudes of transit transmission spectroscopy signals as 2.2$\cdot 10^{-5}$ (b), 3.5$\cdot 10^{-5}$ (c) and 1.2$\cdot 10^{-5}$ (d).

We tested the stability of the system simulating its evolution for $10^{6}$ years with the Mercury package \citep{1999MNRAS.304..793C}, using Bulirsch -- Stoer integrator, adopting circular orbits and masses from the mass - radius relation.
Our simulations show no significant changes in the eccentricity (always below 0.0005 for all the planets) or in the inclination of the orbits (always below $1.6^{\circ}$, $1.2^{\circ}$ and $1^{\circ}$ for planets b, c and d), pointing toward a dynamically stable system.

\section{Conclusions}

We presented a system with three transiting super-Earths orbiting a mid-late type K-dwarf star, discovered with photometric data from K2. The star has been studied and characterized in detail, analyzing its spectrum  and long time photometric series. The detected planets have radii $1.8^{+0.2}_{-0.1}$ $R_{\oplus}$(b), $2.6^{+0.7}_{-0.2}$ $R_{\oplus}$(c), $1.9^{+0.7}_{-0.2}$ $R_{\oplus}$ (c) and orbital periods 
6.342$\pm$0.002 days (b), 13.850$\pm$0.006 days (c), 40.718$\pm$0.005 days (d). 
Additional photometric monitoring is required to confirm planet d and radial velocity monitoring with ultrastable spectrographs at 8-10 m telescopes is necessary to determine accurate planetary  masses. The amplitudes of atmospheric signals in transmission spectroscopy have been estimated at $\sim$ 20 ppm, making the system a good target to incoming facilities such as James Webb Telescope.

\section*{Acknowledgements}
EDA, SLSG, CGG and JCJ acknowledge Spanish ministry projects MINECO AYA2014-57648-P and AYA2017-89121-P, and the regional grant FC-15-GRUPIN14-017. JIGH, BTP and RRL acknowledge the Spanish ministry project MINECO AYA2014- 56359-P. JIGH also acknowledges financial support from the Spanish Ministry of Economy and Competitiveness (MINECO) under the 2013 Ram\'on y Cajal program MINECO RYC-2013-14875. A.S.M acknowledges financial support from the Swiss National Science Foundation (SNSF).





\bsp	
\label{lastpage}
\end{document}